\documentclass[conference,a4paper,10pt]{IEEEtran}

\usepackage[utf8]{inputenc} 
\usepackage[T1]{fontenc}
\usepackage{url}              
\usepackage{cite}             

\usepackage[cmex10]{amsmath}  
\usepackage{mleftright}       
\mleftright                   
\usepackage{amsthm}
\usepackage{graphicx}         
\usepackage{booktabs}         

\usepackage{cite}
\usepackage{amsmath}
\usepackage{amssymb}
\usepackage{algorithmic}
\usepackage{textcomp}
\usepackage{xcolor}
\def\BibTeX{{\rm B\kern-.05em{\sc i\kern-.025em b}\kern-.08em
    T\kern-.1667em\lower.7ex\hbox{E}\kern-.125emX}}
\usepackage{graphicx}
\usepackage{subfiles}
\usepackage[inline]{enumitem}
\usepackage{pifont}

\usepackage{array} 
\usepackage{url}
\usepackage{epstopdf}
\usepackage{tikz}
\usetikzlibrary{automata,arrows,positioning,calc,
shapes,chains}
\usepackage{mathtools}
\usepackage{texdef2015}
\allowdisplaybreaks

\newcommand{\Lcal}{\mathcal{L}}
\newcommand{\Qcal}{\mathcal{Q}}

\newcommand{\R}{\mathbb{R}}

\newcommand{\ql}{q_l}

\newcommand{\laml}[1][l]{\lambda^{(#1)}}

\newcommand{\rowvec}[1]{[\begin{matrix} #1\end{matrix}]}
\newcommand{\rvec}[1]{\smrvec{#1}}
\newcommand{\smrvec}[1]{\setlength\arraycolsep{2pt}\rowvec{#1}}

\newcommand{\piv}{\text{\boldmath{$\pi$}}}

\newcommand{\pibar}{\bar{\pi}}
\newcommand{\vvbar}{\bar{\vv}}

\newcommand{\onev}[1][n]{{\mathbf{1}}_{#1}}
\renewcommand{\vec}[1]{\begin{bmatrix} #1\end{bmatrix}}

\newlength{\swwidth}

\newcommand{\al}{\alpha}

\newcommand{\PR}{r}

\newcommand{\PC}{P_2(\rho)}
\newcommand{\PN}{\Nbar}
\newcommand{\PNn}[1]{\Nbar_{\text{#1}}(\rho)}
\newcommand{\onestar}{1\textsuperscript{*}}
\newcommand{\twostar}{2\textsuperscript{*}}

\newcommand{\MMonestar}{M/M/\onestar}

\usepackage[bottom]{footmisc}

\begin{document}

\title{Timely and Energy-Efficient Multi-Step Update Processing} 

\author{%
  \IEEEauthorblockN{Vishakha Ramani, Ivan Seskar, Roy D. Yates}
  \IEEEauthorblockA{WINLAB,
                   Rutgers University\\
                   Email: \{vishakha, seskar, ryates\}@winlab.rutgers.edu}
}

\maketitle

\begin{abstract}
This work explores systems where source updates require multiple sequential processing steps. We model and analyze the Age of Information (AoI) performance of various system designs under both parallel and series server setups. In parallel setups, each processor executes all computation steps with multiple processors working in parallel, while in series setups, each processor performs a specific step in sequence.
In practice, processing faster is better in terms of age but it also consumes more power. We identify the occurrence of wasted power in these setups, which arises when processing efforts do not lead to a reduction in age. This happens when a fresher update finishes first in parallel servers or when a server preempts processing due to a fresher update from preceding server in series setups.
To address this age-power trade-off, we formulate and solve an optimization problem  to determine the optimal service rates for each processing step under a given power budget. We focus on a special case where updates require two computational steps.

\end{abstract}

\section{Introduction}
Emerging mobile real-time applications such as Augmented Reality (AR) and Mixed Reality (MR) mandate a comprehensive understanding of the surroundings. Sensors supporting such applications generate abundant data that induces a computationally intensive workload not feasible on resource-constrained mobile devices. This challenge is addressed by offloading the computation to a nearby edge node \cite{chen-edge-offloading}.
Examples include personalized AR tours and pedestrian safety systems at smart city intersections \cite{gandhi-2007-pedestrian}, both requiring low latency and timely responses.

Typically, sensor (source) update processing consists of sequential computation steps. For example, the object detection involves sequential tasks like pre-processing on input images, feature extraction, followed by object classification.  
For any edge-computing platform there can be different modes of processing a source update. These modes are usually dictated by system design choices in terms of number of processors deployed to process an update, the underlying communication mechanism between processors as well as energy consumption constraint.

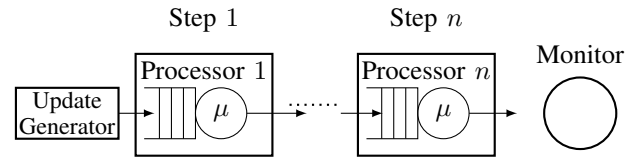
\begin{figure}
\centering
\begin{tikzpicture}[>=latex, scale=0.45]
\draw[thick] (-5,-1.5) rectangle (-2,0);
\node[align=center, font=\small] at (-3.5, -0.75) {Update \\ Generator};
\draw[->] (-2,-0.75) -- +(30pt,0);
\draw[thick] (-1.5,-2) rectangle (2.5,1);
\draw (-1.25,0) -- ++(1.5cm,0) -- ++(0,-1.5cm) -- ++(-1.5cm,0);
\foreach \i in {1,...,3}
  \draw (0.25cm-\i*10pt,0) -- +(0,-1.5cm);
\draw (1,-0.75cm) circle [radius=0.75cm];
\node at (1,-0.75cm) {$\mu$};
\draw[->] (1.75,-0.75) -- +(50pt,0);
\node[align=center] at (0.5cm,0.5cm) {Processor $1$};
\node[align=center] at (0.5cm,2cm) {Step $1$};
\draw[thick, dotted] (3,-0.5) -- (4.5,-0.5);
\draw[->] (4.5,-0.75) -- +(35pt,0);
\draw[thick] (5, -2) rectangle (9,1);
\draw (5.25,0) -- ++(1.5cm,0) -- ++(0,-1.5cm) -- ++(-1.5cm,0);
\foreach \i in {1,...,3}
  \draw (6.75cm-\i*10pt,0) -- +(0,-1.5cm);
\draw (7.5,-0.75cm) circle [radius=0.75cm];
\node at (7.5,-0.75cm) {$\mu$};
\draw[->] (8.25,-0.75) -- +(40pt,0);
\node[align=center] at (7cm,0.5cm) {Processor $n$};
\node[align=center] at (7cm,2cm) {Step $n$};
\draw[thick] (11.5,-0.75) circle (30pt);
\node[align=center] at (11.5cm,1cm) {Monitor};
\end{tikzpicture}
\caption{Processors in series for source update processing. Processor $i$ is responsible for computation step $i$.}
\label{fig:pipeline-cartoon}
\end{figure}

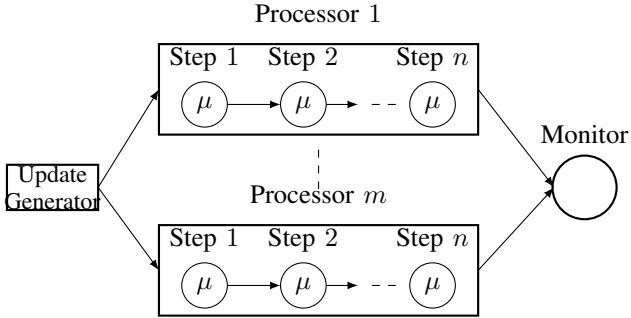
\begin{figure}[t]
\centering
\begin{tikzpicture}[>=latex, scale=0.40]
\draw[thick] (-5,-1.5) rectangle (-2,0);
\node[align=center, font=\small] at (-3.5, -0.75) {Update \\ Generator};

\draw[thick] (0, 1) rectangle (10.5,4);
\node[align=center] at (5.25,5) {Processor $1$};
\draw (1.5,2) circle [radius=0.75cm];
\node[align=center] at (1.5,2) {$\mu$};
\node[align=center] at (1.5,3.5) {Step $1$};
\draw[->] (2.25,2) -- +(50pt,0);
\draw (4.75,2) circle [radius=0.75cm];
\node[align=center] at (4.75,2) {$\mu$};
\node[align=center] at (4.74,3.5) {Step $2$};
\draw[->] (5.5,2) -- +(30pt,0);
\draw[dashed] (7,2) -- (8,2);
\draw (9,2) circle [radius=0.75cm];
\node[align=center] at (9,2) {$\mu$};
\node[align=center] at (9,3.5) {Step $n$};

\draw[dashed] (5.25, 0.5) -- (5.25,-0.8);

\draw[thick] (0, -5) rectangle (10.5,-2);
\node[align=center] at (5.25,-1) {Processor $m$};
\draw (1.5,-4) circle [radius=0.75cm];
\node[align=center] at (1.5,-4) {$\mu$};
\node[align=center] at (1.5,-2.5) {Step $1$};
\draw[->] (2.25,-4) -- +(50pt,0);
\draw (4.75,-4) circle [radius=0.75cm];
\node[align=center] at (4.75,-4) {$\mu$};
\node[align=center] at (4.74,-2.5) {Step $2$};
\draw[->] (5.5,-4) -- +(30pt,0);
\draw[dashed] (7,-4) -- (8,-4);
\draw (9,-4) circle [radius=0.75cm];
\node[align=center] at (9,-4) {$\mu$};
\node[align=center] at (9,-2.5) {Step $n$};

\draw[->] (-2, -0.75) -- (0,2.5);
\draw[->] (-2, -0.75) -- (0, -3.5);
\draw[thick] (14,-0.75) circle (30pt);
\node[align=center] at (14cm,1cm) {Monitor};
\draw[->] (10.5, 2.5) -- (13,-0.75);
\draw[->] (10.5, -3.5) -- (13, -0.75);
\end{tikzpicture}
\caption{Parallel processors setup for source update processing. Each processor executes all $n$ computation steps. }
\label{fig:parallel-cartoon}
\end{figure}

One approach involves using $n$ loosely coupled processors to process an update requiring $n$ computational steps. In this configuration, each processor performs one step in the update's processing pipeline. This setup introduces an asynchronous pipeline mechanism, where the output of a processor serves as the input to the subsequent processor. From a queueing theory perspective, this can be modeled as a tandem queue (also known as a series queue) with $n$ servers (as depicted in Fig.~\ref{fig:pipeline-cartoon}). 
In contrast, another processing paradigm involves the use of multiple parallel processors, where each processor independently executes all $n$ computation steps, as illustrated in Fig.~\ref{fig:parallel-cartoon}.


\subsection{AoI and  Multi-Step Processing Systems}
Regardless of the mode of processing, it is imperative that the system 
delivers timely processed updates such that the age at the end user, hereafter referred to as {\em monitor}, is minimized. 
Despite its importance, the study of timely multi-step update processing remains unexplored in AoI literature. 
This work aims to bridge that gap by investigating the age performance of two-step ($n=2$) update processing systems, a fundamental building block for more complex processing systems.
Even within this seemingly simple two-step framework, rudimentary questions arise. For instance, which configuration-—series or parallel processing—-proves more effective in maintaining timeliness? Answering this question, however, is far from straightforward and poses a considerable challenge.

In a setup with two servers in series, modeled as a tandem queue, each service facility may operate under different service disciplines, such as lossless First-Come-First-Served (FCFS) or lossy Last-Come-First-Served (LCFS) with preemption in either waiting or service steps. 
While there is an extensive literature on age performance of fresh arrivals in single-source single-server queues employing various service disciplines (see \cite{Yates-SBKMU-2021jsac-survey} and the references therein), in our work however, updates arrive at server $2$ from server $1$ with some existing age, which must be accounted for by the analysis.

Furthermore, the analysis of  parallel processor setups with  only two servers is equally non-trivial. A server may be ``late'' in delivering an update, while another parallel server has already delivered an update with lesser age, rendering the former's delivery inconsequential in terms of age reduction. This scenario underscores the necessity of developing a novel analytical framework to properly evaluate such systems.

\subsection{Age-Power Trade-off}
We argue that it is not straightforward to determine which setup -- series or parallel servers -- performs better solely based on age. 
While faster processing improves age performance, it also consumes more power.
A deeper analysis reveals that both configurations are susceptible to a phenomenon we term  {\em wasted power}, where computational resources are expended on processing updates that ultimately do not contribute to reducing the age.

In parallel server setups, wasted power occurs when one server completes processing a fresher update before others, thereby rendering the subsequent efforts of the remaining servers, still processing older updates, inconsequential in terms of age reduction. The resources dedicated to these outdated updates are thus wasted.
In series server setups, wasted power may occur when a server preempts its current update upon receiving a fresher update from the preceding server. 
The computational resources expended in processing the preempted update can be identified as wasted. 
Similarly, in some scenarios, a server may discard updates received from the preceding server, nullifying the work performed on discarded update.

An additional inefficiency in series setups is server idleness. Servers may remain inactive while awaiting the completion of the preceding server’s processing task. This idle time represents lost computational potential, as these resources could have been employed to process other updates, potentially improving age performance.

\subsection{Paper Outline and Main Contributions}
In this work, we address the age-power trade-off and pose the following question: How can we optimize the allocation of computational resources to minimize wasted power while simultaneously reducing age? We explore this trade-off by focusing primarily on identifying the optimal service rates for each step that minimize age when the system is subjected to a total power consumption constraint. We start with
Section~\ref{sec:system-model} which introduces the system parameters, the power consumption model, and the optimization problem formulation. 

We focus on series server setups in Section~\ref{sec:seq-servers-models}. Specifically, we analyze the power consumption and age performance of various preemptive and non-preemptive tandem queue models. These models differ in the queuing discipline employed at server $2$, while server $1$ generates updates at will.
Particularly, we study four models under series server setup: \MMonestar, M/M/1/\twostar, M/M/1/1, Synchronous Sequential Service (SSS). In the \MMonestar{} model, server $2$ employs preemption in service, discarding the current update upon the arrival of a new one. In the M/M/1/\twostar{} model, server $2$ includes a waiting room of capacity 1, where updates in the waiting room can be preempted by new arrivals.
In the M/M/1/1 model, server $2$ blocks and discards incoming updates when it is busy.
In SSS model, a new update begins service only when both servers are idle.

In Section~\ref{sec:parallel-serv-models}, our focus turns to parallel server setups. 
We begin by analyzing the power consumption and age in the baseline model, Parallel Synchronous Sequential Service (P-SSS), where the two servers function independently. We introduce a novel approach using the Stochastic Hybrid Systems (SHS) methodology to derive a system of linear equations for calculating the average age at the monitor.
Subsequently, we analyze power consumption and age performance of two additional policies, Parallel Coordinated Alternating Freshness (P-CAF) and Parallel Shared Intermediate Update (P-SIU), where servers synchronize their operations by leveraging information about each server's current step of processing. 

In Section~\ref{sec:num-eval},
we numerically solve the optimization problem for various series and parallel server models introduced in Sections~\ref{sec:seq-servers-models} and \ref{sec:parallel-serv-models}. This includes a comparative analysis of the minimum achievable age for each model, obtained by optimizing the computational step rates subject to a given power constraint. 
In Section~\ref{sec:open-probs}, we conclude by discussing several open problems emerging from this work and propose directions for future research.

\subsection{Related Work}
In the Age-of-Information literature, various studies have focused on age in network of queues \cite{Bedewy-SS-isit2017, Talak-KM-allerton2017, Yates-aoi2018}. For a line network model of last-come-first served (LCFS) queue with preemption in service,  it was shown that node $i$ with service rate $\mu_i$ contributes $1/\mu_i$ to the age at the monitor~\cite{Yates-aoi2018}. Authors in \cite{Kam-KNWE-milcom2017} derived average age for two first-come-first served non-preemptive queues in tandem. \cite{Chiariotti-toc2021} models the communication and computation delay in edge computing framework and derives the PDF of Peak Age-of-Information (PAoI) for M/M/1-M/D/1 and M/M/1-M/M/1 tandem queues. \cite{sinha2024-tandem-paoi} develops a recursive framework to derive the mean peak age of information for $N$ heterogeneous servers in tandem.
\cite{popovski-tandempriority-globecom2024} obtains the distribution of the age and peak age in a system of two tandem queues connected in series with packet prioritization in the second queue.

Age for M/M/2 and M/M/$\infty$ systems was studied in \cite{Kam-KNE-IT2016diversity} to demonstrate the advantage of having the message transmission path diversity for status updates. \cite{TalakTransIT2021} studies the age-delay trade-off in G/G/$\infty$ queue.
\cite{Fidler-JSAC-mm1parallel-2024} observes that a single M/M/1 queue has better age performance than the independent parallel M/M/1 queues with the same total capacity.
\cite{Yates-isit2018} analyzed age in network of parallel finite identical and memoryless servers, where each server is an LCFS queue with preemption in service. However, our work deviates from \cite{Kam-KNE-IT2016diversity, Yates-isit2018} in that we relax the assumption of memoryless processing times for updates. This key difference renders the SHS analysis used in \cite{Yates-isit2018} inapplicable to our scenario.

On the other hand, with respect to general queuing theory,
the problem of optimal service rate control has been extensively studied across various types of queuing networks, ranging from single-queue single server model \cite{George-sing-q-single-serv, crabill-sing-q-single-serv, stidham-sing-q-sing-serv}, multiple queue single server model \cite{Hofri-multi-q-single-serv}, to multiple server, multiple queue model \cite{lee-kulkarni-multiserv-2014}.
In studies focused on single-server queue systems, the general setting involves a nondecreasing cost of service and holding costs that are nondecreasing functions of queue length, with rewards associated with customers entering the queue. The arrival rate, $\lambda$, and/or the service rate, $\mu$, are subject to control. The objective in these studies is typically to minimize the expected total discounted cost or the long-run average cost. 
In various systems, the authors establish optimality of monotone policies i.e. optimal arrival rates are non-increasing in number of arrivals and optimal service rates are non-decreasing in queue length as observed in \cite{weber-cyclicqueue}. 

Several authors have considered tandem queue systems with Poisson arrivals at rate $\lambda$ and two memoryless servers, serving at rates $\mu_1$ and $\mu_2$ at first and second queue respectively. The first study on optimal service control in tandem queues was conducted by Rosberg et al. \cite{rosberg-tandem}. In this study, the authors examined a setting where the service rate at server $1$ is selected as a function of the system's state, defined as the tuple of queue lengths at each server, while the service rate at server $2$ is held constant. Considering only holding cost and no operating cost, the authors established the optimality of switchover policies, where the optimal rate at server $1$ is determined by a switching function of the queue length at server $2$.

Authors in \cite{weber-cyclicqueue} considered a cyclic queue system where a number of $\cdot$/M/1 queues are arranged in a cycle. 
Considering a system cost comprising of both holding and operating costs, the authors determined the optimal policy has a transition-monotone decision rule, where when a customer moves from queue $i$ to the following queue, the optimal service rate at queue $i$ does not increase, and optimal service rate at queue $j, j\neq i$ does not decrease.
Optimal control of service rates of a tandem queue under power constraints is studied in \cite{xia-tandem-powercons}. The authors assume that the service rate is linear to the power allocated to that server and the sum of service rates must not exceed the given power budget. An iterative algorithm is proposed to find the optimal service rates.

\section{System Model Overview}
\label{sec:system-model}
\subsection{Power Consumption Model}
Power dissipation in digital CMOS circuits is primarily attributed to dynamic power, short circuit losses, and transistor leakage currents \cite{Chandrakasan-CMOS}. Among these, dynamic power consumption is currently the main component in high-performance microprocessors. Dynamic power, driven by the periodic switching of capacitors, can be approximated by the well-known formula
\begin{equation} \eqnlabel{cmos-power}
    P = A C_L V^2 f,
\end{equation}
where $A$ and $C_L$ denote the Activity Factor (AF) and loading capacitance, respectively, $V$ is the supply voltage, and $f$ is the clock frequency \cite{Magen-powerdissipation-2004}.
According to alpha-power law MOS model \cite{Sakurai-Newton-JSSC},
$f \propto V^{\al_c-1}$, where $\al_c$, called the velocity
saturation index, is a technology dependent factor, typically ranging between $1$ and $2$. This implies that voltage can be expressed as 
$V = k_V f^{1/(\al_c-1)}$, with $k_V$ denoting a constant of proportionality.
This results in a power model for a processor running at frequency $f$ as
\begin{equation} \eqnlabel{power-freq}
    P = k_P f^\al,
\end{equation}
where $\al = (1+\al_c)/(\al_c-1) \ge 3$ and $k_P = A C_L k_V^2$ is a constant that incorporates the constants $A$ and $C_L$ from \eqnref{cmos-power}.
According to \cite{Gonzalez-JSSC1997cmos}, for a 25 $\mu$m technology, $\al_c$ is likely to be in range $\rvec{1.3,1.5}$.
For our numerical evaluations,  we fix the velocity saturation index at $\al_c = 1.5$, which corresponds to $\al = 5$.

\subsection{Processing Speed and Power Consumption}
We assume that processing a source update involves a sequence of two computational steps. 
Further, we consider that the system uses two servers, which may operate in series or in parallel configurations.  

In general, each step $i$ in update processing involves a random computation workload $C_i$, measured in CPU cycles.
This randomness is due to many reasons such as  CPU being shared by many applications, background daemons, garbage collection and other system-level activities \cite{dean-tailatscale}. While ideally,
the workload in terms of CPU cycles required to process the same update would remain consistent across servers, variations occur due to these random factors.

For simplicity, we assume that the workloads for the two steps are identically distributed, with means $\E{C_1} = \E{C_2} = \E{C}$. Each step $i$ is executed at a constant processing frequency $f_i$ (CPU cycles per unit time), resulting in an execution time $T_i = C_i/f_i$. The average service rate for step $i$ is therefore 
\begin{equation}
    \mu_i = \frac{1}{\E{T_i}}=\frac{f_i}{\E{C_i}} = \frac{f_i}{\E{C}}.
\end{equation} 

While executing step $i$, the processor consumes power as described by \eqnref{power-freq}, resulting in instantaneous power consumption: $P_i =  k_Pf_i^\al =  k_P(\E{C}\mu_i)^\al$.  
To simplify the relation between power consumption and processor speed, we assume a power unit such that $k_P=1$. Thus, in our analysis, the expected power consumed by a processor while executing step $i$ simplifies to $P_i=(\E{C} \mu_i)^\al$. 
Finally, we assume that an idle processor consumes negligible power.

We assume that the execution time for step $i$ follows an independent exponential distribution with rate parameter $\mu_i$,
i.e., $T_i \sim \exp(\mu_i)$.
Let $\Qcal$ represent the discrete state space of a given update processing system. That is, a state $q\in\Qcal$ specifies for each processor whether it is idle or executing update step $1$ or step $2$.  
With $\pi_q$ representing the stationary probability of state $q \in \Qcal$,
the power consumption associated with step $i$ execution in state $q$ is represented as $\PR_{q,i}(\mu_i)$. For example, if in state $q$, both processors are executing step $1$ for a particular update, $\PR_{q,1}(\mu_1)=2\E{C}^\alpha\mu_1^\alpha$ while  $\PR_{q,2}(\mu_2)=0$ because no processor is executing step $2$. In general, by defining $n_{q,i}$ as the number of processors working on step $i$ in state $q$, 
\begin{equation}
\PR_{q,i}(\mu_i)=n_{q,i}(\E{C}\mu_i)^\alpha
\eqnlabel{PRcount}
\end{equation}
The average power consumption associated with step $i$ is then $\sum_{q \in \Qcal} \pi_q \PR_{q,i}(\mu_i)$. 

\subsection{Problem Formulation}
We enforce that the total power consumption at the two servers is limited by a power budget. 
Specifically, with $P$ representing the total power budget and  $\pi_q(\mu_1, \mu_2)$ elucidating the dependence of $\pi_q$ on $\mu_1$ and $\mu_2$, the power consumption at the two servers must satisfy the constraint
\begin{equation}
\sum_{q \in \Qcal} \pi_q(\mu_1, \mu_2)[ r_{q,1}(\mu_1) + r_{q,2}(\mu_2)] \le P. \eqnlabel{power-constraint-2}
\end{equation}
The age at the monitor, denoted by $\Delta(\mu_1,\mu_2)$,  is a function of the service rates $\mu_1$ and $\mu_2$, and is
influenced by the policy at each server. 
Our objective is to minimize the age $\Delta(\mu_1,\mu_2)$ at the monitor by controlling the service rates $\mu_1$ and $\mu_2$, subject to the power constraint \eqnref{power-constraint-2}. From \eqnref{PRcount} and \eqnref{power-constraint-2},
the optimization problem can be stated as:
\begin{subequations} \eqnlabel{orig-opt-prob}
\begin{alignat}{2}
    & \text{minimize} &\quad& \Delta(\mu_1, \mu_2) \\
    & \text{subject to} & &  
    \sum_{q \in \Qcal} \pi_q(\mu_1, \mu_2) \sum_{i=1}^2n_{q,i}\mu_i^\alpha 
\le P/\E{C}^\alpha, \eqnlabel{power-constraint-3} \\
    &&& \mu_1, \mu_2 \ge 0.
\end{alignat}
\end{subequations}

To solve the 
optimization problem \eqnref{orig-opt-prob}, our strategy is to define
$\rho = \mu_1/\mu_2$, and focus on a class of systems in which the stationary probabilities $\pi_q$ can be expressed as functions of $\rho$ i.e., 
$\pi_q(\mu_1,\mu_2) \equiv \pi_q(\rho)$.
In 
the sequential service system of Figure~\ref{fig:pipeline-cartoon}, $\rho$ represents the total offered load from server $1$ to server $2$. In general, $\rho$ characterizes the  effort that processors make on step $1$ relative to step $2$

Substituting $\mu_1 = \rho \mu_2 $ into the constraint \eqnref{power-constraint-3} results in
\begin{equation} \eqnlabel{power-constraint}
\sum_{q \in \Qcal} \pi_q(\rho) (n_{q,1}\rho^\alpha\mu_2^\alpha + 
n_{q,2}\mu_2^\alpha) \le P/\E{C}^\alpha.
\end{equation}
With the observation that
\begin{equation}
\Nbar_i(\rho) =\sum_{q\in\Qcal} \pi_q(\rho) n_{q,i}
\end{equation}
is the average number of processors working on step $i$,  
we see that \eqnref{power-constraint} simplifies to the upper bound on the step 2 service rate
\begin{equation}
    \mu_2^\al \le \frac{P}{\E{C}^\al(\rho^\alpha\Nbar_1(\rho)+\Nbar_2(\rho))}.
\end{equation}
Defining the Power-Weighted Processor Activity (PWPA) 
\begin{align}
\PN(\rho)&\equiv\rho^\al \Nbar_1(\rho) +\Nbar_2(\rho), \eqnlabel{pn-def}
\end{align}
the optimization problem \eqnref{orig-opt-prob} can be reformulated as
\begin{subequations} \eqnlabel{opt-problem}
\begin{alignat}{2}
    & \text{minimize} &\quad& \Delta(\mu_2, \rho) \\
    & \text{subject to} && \mu_2^\al \le  \frac{P}{\E{C}^\al\Nbar (\rho)}.
\eqnlabel{mu2-constraint} \\
    &&& \mu_2 \ge 0, \text{and } \rho \ge 0.
\end{alignat}
\end{subequations}

The parameter $\al$ is fixed by the technology, while the power budget $P$ and CPU demand $C$ are system parameters. 
For a fixed $\rho$, the service rates $\mu_2$ and $\mu_1=\rho\mu_2$ determine the average age at the monitor. 
We will see in the systems we study that the age is typically minimized by choosing $\mu_2$ as large as possible subject to the upper bound  \eqnref{mu2-constraint}.
What remains is choosing the  right value of $\rho$. 
A larger $\rho$ implies that step 1 is executed faster relative to step 2, resulting in more energy being allocated to step 1.
This results in fresher updates reaching step 2, but
this could be wasting the energy used in step 1, as updates from step 1 could either be discarded or updates in step 2 could be preempted. 
On the other hand, a smaller $\rho$ means that step 1 is slower relative to step 2. In this case, step 1 processed updates arrive at step 2 with higher age with the system not feeding enough updates to step 2. 


For a system design choice, 
finding an optimal $\rho^*$ allows us to determine the corresponding optimal service rate $\mu_2^*$ using the constraint in \eqnref{mu2-constraint}, which in turn yields the optimal age $\Delta(\mu_2^*, \rho^*)$. 
The subsequent sections analyze the system models studied in this work. Specifically, we detail the relevant Markov Chain for each model, including its state space $\Qcal$, stationary probabilities $\pi_q$, and transitions. Analytical expressions for $\Delta(\mu_2, \rho)$ and $\PN(\rho)$ 
are provided for each model, enabling the formulation of optimization problem \eqnref{opt-problem} for each model.

\subsection{SHS Overview}
We use Stochastic Hybrid Systems (SHS)~\cite{hespanha2006modelling}
to evaluate AoI of processed updates. The SHS based approach for AoI evaluation was first introduced in \cite{yates2018ToIT} and has been since employed in AoI evaluation of a variety of status updating systems \cite{Yates-aoi2018,Farazi-KB-aoi2018,Maatouk-AE-aoi2019,Kaul-Yates-isit2018priority,Maatouk-AE-ToN2020,Yates-IT2020,Moltafet-LC-CommLetters2021,Moltafet-LC-ISWCS2021}.
An SHS has a state-space with two components -- a  discrete component $q(t) \in \Qcal = \{0, 2, \ldots, M\}$ that is a continuous-time finite-state Markov Chain and a continuous component $\xv(t) = [x_0(t), \ldots , x_n(t)] \in \R^{n+1}$. In AoI analyses using SHS, each $x_{j}(t) \in \xv(t)$ describes an age process of interest. Each transition $l \in \Lcal$ is a directed edge $(q_l, q'_l)$ with a transition rate $\laml$ in the Markov chain. The age process vector evolves at a unit rate in each discrete state $q \in \Qcal$, i.e., $\frac{d\xv}{dt} = \dot{\xv}(t) = \onev$. A transition $l$ causes a system to jump from discrete state $q_l$ to $q_l^\prime$ and resets the continuous state from $\xv$ to $\xv'$ using a linear transition reset map $\Amat_l \in \{0,1\}^{(n \times n)}$ such that $\xv' = \xv \Amat_l$. For simple queues, examples of transition reset mappings $\set{\Amat_l}$  can be found in \cite{yates2018ToIT}.

For a discrete state $\qbar \in \Qcal$, let 
\begin{align}
    \Lcal_{\qbar} &= \{l \in \Lcal : q'_l = \qbar \}, &
    \Lcal'_{\qbar} &= \{l \in \Lcal : q_l = \qbar \}.
\end{align}
denote the respective sets of incoming and outgoing transitions.
Age analysis using SHS is based on the  expected value processes $\set{\vv_q(t)\colon q\in \Qcal}$ such that 
 $\vv_q(t)
 =\E{\xv(t)\delta_{q,q(t)}}$,
 with $\delta_{i,j}$ denoting the Kronecker delta function. For the SHS models of age processes considered here, each $\vv_q(t)$ will converge to a fixed point $\vvbar_q$. The fixed points $\set{\vvbar_q\colon q\in\Qcal}$ are the solution to a set of age balance equations. The following theorem provides a simple way to calculate the age-balance fixed point and then the average age. 
\begin{theorem}\thmlabel{AOI-SHS}
\cite[Theorem~4]{yates2018ToIT}
If the discrete-state Markov chain $q(t)\in\Qcal=\set{0,\ldots,M}$ is ergodic with stationary distribution 
$\bar{\piv}=\rvec{\bar{\pi}_0 &\cdots &\bar{\pi}_M}>0$ and there exists a 
non-negative vector $\vvbar=\rvec{\vvbar_0&\cdots&\vvbar_M}$
such that 
\begin{align}
\bar{\vv}_{\qbar}\sum_{l\in\Lcal_{\qbar}}\laml &=\onev[]\bar{\pi}_{\qbar}+ \sum_{l\in\Lcal'_{\qbar}}\laml \bar{\vv}_{\ql}\Amat_l,\quad \qbar\in\Qcal,\eqnlabel{AOI-SHS-v}
\end{align}
then
the average age vector is 
$\E{\xv}\!=\!{\displaystyle \limty{t}}\E{\xv(t)}\!=\!
\sum_{\qbar\in\Qcal} \vvbar_{\qbar}$.
\end{theorem}

\section{Problem Formulation: Servers in Series} \label{sec:seq-servers-models}
In this section, we analyze update processing models with two servers arranged in series, as depicted in Fig. \ref{fig:pipeline-cartoon} for the case of $n=2$. Server 1 (Processor 1) performs the first step of processing, while server 2 (Processor 2) handles the second step.
We assume a generate-at-will with zero-wait scenario at server $1$ such that it can generate a fresh (age zero) update whenever it wishes.  
However, we consider variations on service disciplines at server $2$. 
There may be a single queue to save updates from
server $1$ when server $2$ is busy. 
Since, the queuing (if any) is only at server $2$, we name our sub-models based on the queuing discipline at server $2$. 

With generate-at-will with zero-wait strategy and memoryless service times, the departure process at server 1 is a Poisson process with rate $\mu_1$. Consequently, the inter-arrival times of updates at server 2 follow an exponential distribution with parameter $\mu_1$. The service time at server $2$ is also exponential, with rate $\mu_2$.

We adopt Kendall's notation to denote the queuing discipline at server $2$, following the convention used in the AoI literature \cite{Costa-CE-IT2016management, Yates-SBKMU-2021jsac-survey}.  
For example, an \text{M/M/1/1} submodel implies a queueing system that blocks and clears a new arrival while server $2$ is busy.
We use the notation \MMonestar{} to indicate preemption in service at server $2$, and M/M/1/\twostar{} to denote a system with a waiting room having an update capacity of 1, with preemption in waiting. 
We now describe the analysis of various preemptive and non-preemptive system models for servers in series configuration.

\subsection{\MMonestar}
In this model, server $1$ generates a fresh update immediately upon completing the processing of the previous update. 
The update is then passed to server $2$ at a rate $\mu_1$.
Server $2$ employs preemption in service, allowing a new arrival from server $1$ to preempt an update currently being serviced at server $2$. Consequently, an update departing from server $1$ immediately enters service at server $2$, and any preempted update at server $2$ is discarded. Since there is no queuing at server $2$, it is either idle or actively serving an update. 
The discrete state space of \MMonestar{} is $\Qcal=\{0,1\}$ where $0$ corresponds to server $2$ being idle and state $1$ represents busy server $2$. The stationary probabilities are
\begin{equation}
   \pi_0 =  \frac{1}{1+\rho}, \quad\text{and} \quad \pi_1 = \frac{\rho}{1+\rho}.
    \eqnlabel{piq-mm1s}
\end{equation}
Our \MMonestar{} model is analogous to the line network studied in \cite{Yates-aoi2018, Yates-IT2020}, where it was demonstrated that the age at the monitor for a two-server line network, applicable to our model as well, is given by
\begin{equation}
\Delta_\text{\MMonestar}(\mu_1, \mu_2) = \frac{1}{\mu_1} + \frac{1}{\mu_2}, \eqnlabel{mm1s-age-mu1mu2}
\end{equation}
Alternatively, we can express the age in terms of $\mu_2$ and $\rho$ as
\begin{equation}
    \Delta_\text{\MMonestar}(\mu_2,\rho) = \frac{1}{\mu_2}\left(1+ \frac{1}{\rho}\right). \eqnlabel{mm1s-age-mu2rho}
\end{equation}
Since server $1$ is perpetually busy with step 1, $\Nbar_1(\rho) =1$.
Since server 2 works on step 2 only in state $1$, \eqnref{piq-mm1s} implies
\begin{equation} \eqnlabel{n2bar-mm1s}
    \Nbar_2(\rho) = \sum_{q\in\Qcal}\pi_q(\rho) n_{q,2} = \pi_1(\rho)=\frac{\rho}{1+\rho}.
\end{equation}
It then follows from \eqnref{pn-def} 
that the power-weighted processor activity $\PN(\rho)$ for \MMonestar{} 
is 
\begin{equation} \eqnlabel{pn-mm1s}
    \PNn{\MMonestar} = \rho^\al\Nbar_1(\rho) + \Nbar_2(\rho) = \rho^\al + \frac{\rho}{1+\rho}.
\end{equation} 
Thus, the \MMonestar{} speed constraint  \eqnref{mu2-constraint} takes the form 
\begin{equation} \eqnlabel{mu2-const-mm1s}
    \mu_2^\al \le \frac{P}{\E{C}^\al\PNn{\MMonestar}}.
\end{equation}

\subsection{M/M/1/2*}
Server $1$ generates a fresh update as soon as it finishes processing the previous update. 
The step $1$ update is then sent to the waiting room of server $2$, which has a capacity of $1$. In this waiting room, a new arrival from server $1$ preempts any existing update. 
If the waiting room is empty, an arrival from server 1 goes into service at server 2.
When server $2$ completes processing its current update, it either sits idle if the waiting room is empty or begins processing the next update from the waiting room. 
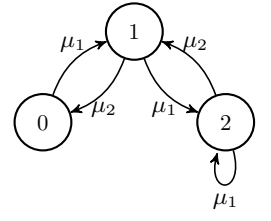
\begin{figure}[t]
$\setlength{\extrarowheight}{0.5mm}
\begin{array}[b]{cccc}
l & q_l\to q^\prime_l & \laml &  \xv\Amat_l \\ \hline
1 & 0\to1 & \mu_1 & \rvec{x_0, 0, x_1, x_1} \\
2 & 1\to0 & \mu_2 & \rvec{x_2, x_1, x_2, x_3} \\
3 & 1\to2 & \mu_1 & \rvec{x_0, 0, x_2, x_1} \\
4 & 2\to2 & \mu_1 & \rvec{x_0, 0, x_2, x_1} \\
5 & 2\to1 & \mu_2 & \rvec{x_2, x_1, x_3, x_3}
\end{array}$
\hfill
\begin{tikzpicture}[->, >=stealth', auto, semithick, node distance=2cm]
\tikzstyle{every state}=[fill=none,draw=black,thick,text=black,scale=0.85]
\node[state] (1)                    {$1$};
\node[state] (0) [below left of=1]  {$0$};
\node[state] (2) [below right of=1] {$2$};
\path
(0) edge[bend left=25, above] node{\small $\mu_1$}  (1)
(1) edge[bend left=25, below] node{\small $\mu_2$} (0)
(1) edge[bend right=25, below] node{\small $\mu_1$} (2)
(2) edge[bend right = 25,above] node{\small $\mu_2$} (1)
(2) edge[loop below] node{\small $\mu_1$} (2);
\end{tikzpicture}
\caption{The SHS transition/reset maps and Markov chain corresponding to M/M/$1$/$2^*$ model.}
\label{fig:mm12s-shs}
\end{figure}
The age of processed update can be analysed using the SHS Markov chain and table of state transitions depicted in Fig.~\ref{fig:mm12s-shs}. 
The continuous state age vector is $\xv = \rvec{x_0,x_1, x_2, x_3}$, where 
$x_0$ is the age of the processed update at the monitor,
$x_1$ and $x_2$ are the ages of the update at server $1$ and server $2$ respectively,
and $x_3$ is the age of the update at server $2$'s waiting room.
The discrete state is $\Qcal = \set{0,1,2}$, where state $0$ corresponds to server $2$ being idle, and states $1$ and $2$ correspond to server $2$ being busy with no update in the queue and one update waiting in the queue, respectively.

We now describe SHS transitions enumerated in the table in Fig.~\ref{fig:mm12s-shs}. 
\begin{itemize}
\item {$l=1$:} Server $1$ finishes step $1$ and sends the update to idle server $2$. Server $2$ receives an update of age $x_1$, thus $x_2' = x_1$. 
A fresh update is generated at server $1$, thus $x_1' = 0$. Age at the monitor remains unchanged, hence $x_0' = x_0$.
\item {$l=2$:} Server $2$ finishes step $2$, and delivers the update to monitor, making $x_0'=x_2$. The waiting room is empty, and server $2$ waits for an update from server $1$, resulting in no change in $x_2$.
\item {$l=3,4$:} 
Update from server $1$  arrives in the waiting room and preempts the update (if any), resetting the age in the waiting room to $x_3' = x_1$. Server $1$ generates a fresh update, hence $x_1' = 0$. 
\item {$l=5$:} Server $2$ finishes step $2$ and delivers update to the monitor, resulting in $x_0' = x_2$. Since there is an update waiting in server $2$'s buffer with age $x_3$, server $2$ starts processing this update, thus age at server $2$ is reset to the age of update in the waiting room i.e., $x_2'=x_3$.
\end{itemize}

The Markov chain in Fig.~\ref{fig:mm12s-shs} has stationary probabilities $\piv$
with normalization constant $C_\pi$ given by
\begin{subequations}
\begin{align}
    \piv &=  \rvec{\pi_{0} & \pi_1 & \pi_2}
    = C_\pi^{-1} \rvec{1 & \rho & \rho^2},\\
C_\pi &= 1+\rho+\rho^2.
\end{align}
\end{subequations}
We now use \Thmref{AOI-SHS} to solve for
$\vvbar = \rvec{\vvbar_{0}&\vvbar_{1}&\vvbar_{2}}$,
where 
$\vv_q=\rvec{v_{q0}&v_{q1}&v_{q2}&v_{q3}}, \forall q \in \Qcal$. 
With $\mu_1=\rho\mu_2$, this yields
\begin{subequations}
\eqnlabel{mm12s-shs-eqn}
\begin{align}
\rho\mu_2 \vvbar_0 &= \onev[]\pibar_0 + \mu_2\vvbar_1\Amat_2, \\
\mu_2(1+\rho) \vvbar_1 &= \onev[]\pibar_1 + \rho\mu_2 \vvbar_0\Amat_1 + \mu_2\vvbar_2\Amat_5, \\
\mu_2(1+\rho)\vvbar_2 &= \onev[]\pibar_2 + \rho\mu_2\vvbar_2\Amat_4 + \rho\mu_2\vvbar_1\Amat_3.
\end{align}
\end{subequations}
The age at the monitor $\Delta_\text{M/M/1/\twostar}$,  
is then calculated as $\Delta_\text{M/M/1/\twostar}= v_{0,0} + v_{1,0} + v_{2,0}$. Some algebra yields\footnote{Note that the M/M/1/\twostar{}
 model  here is equivalent to the  end-to-end update processing  model  in the edge computing scenario of\cite{gong2019-wcsp}. With arrival rate $\lambda=\mu_1$ and service rate $\mu=\mu_2$, the age expression \cite[Theorem 1, Equation (9)]{gong2019-wcsp} derived with sawtooth waveform analysis can be shown to be identical to 
\eqnref{mm12s-age-mu2rho}.}
\begin{align}
&\Delta_\text{M/M/1/\twostar}(\mu_2, \rho)\nn 
&\quad=
\frac{1}{\mu_2}\Bigl(\frac{2}{\rho} + \frac{2\rho^2}{1+\rho+\rho^2} + \frac{(1+2\rho)(1+3\rho+\rho^2)}{(1+\rho)^4}\Bigr).
\eqnlabel{mm12s-age-mu2rho}
\end{align}
Similar to \MMonestar, server 1 is always busy with step 1. As such 
for each state $q \in \Qcal$, $n_{q,1} = 1$. The average number of processors working on step 1 is
\begin{equation} \eqnlabel{n1bar-mm12s}
    \Nbar_1(\rho) = \sum_{q\in\Qcal}\pi_q(\rho) n_{q,1} = 1.
\end{equation}
Only server 2 works on step 2 in states $q=1$ and $q=2$. No server works on step 2 in state $q=0$. Hence,
\begin{equation} \eqnlabel{n2bar-mm12s}
    \Nbar_2(\rho) = \sum_{q\in\Qcal}\pi_q(\rho) n_{q,2} = \pi_1 + \pi_2 = \frac{\rho(1+\rho)}{1+\rho+\rho^2}.
\end{equation}
From \eqnref{pn-def}, \eqnref{n1bar-mm12s} and \eqnref{n2bar-mm12s}, the power-weighted processor activity $\PN(\rho)$ takes the form
\begin{equation} \eqnlabel{pn-mm12s}
\PNn{M/M/1/\twostar} = \rho^\al\Nbar_1(\rho) + \Nbar_2(\rho) = \rho^\al + \frac{\rho(1+\rho)}{(1+\rho+\rho^2)}.
\end{equation}
For M/M/1/\twostar, \eqnref{mu2-constraint} and \eqnref{pn-mm12s} imply 
\begin{equation}
    \mu_2^\al \le \frac{P}{\E{C}^\al\PNn{M/M/1/\twostar}}.
\end{equation}


\begin{figure}[t]
$\setlength{\extrarowheight}{0.5mm}
\begin{array}[b]{cccc}
l & q_l\to q^\prime_l & \laml &  \xv\Amat_l \\ \hline
1 & 0\to1 & \mu_1 & \rvec{x_0, 0, x_1} \\
2 & 1\to1 & \mu_1 & \rvec{x_0, 0, x_2} \\
3 & 1\to0 & \mu_2 & \rvec{x_2, x_1,x_2}
\end{array}$
\hfill
\begin{tikzpicture}[->, >=stealth', auto, semithick, node distance=1.8cm]
\tikzstyle{every state}=[fill=none,draw=black,thick,text=black,scale=0.85]
\node[state] (0)                    {$0$};
\node[state] (1) [right of=0]  {$1$};
\path
(0) edge[bend left=25, above] node{\small $\mu_1$}  (1)
(1) edge[bend left=25, below] node{\small $\mu_2$} (0)
(1) edge[loop right] node{\small $\mu_1$} (1);
\end{tikzpicture}
\caption{The SHS transition/reset maps and Markov chain corresponding to Model $M/M/1/1$.}
\label{fig:model-3-shs}
\end{figure}
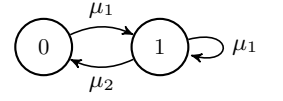
\subsection{M/M/1/1}
In this model as well, server $1$ generates a fresh source update immediately after processing the previous one. Server $1$ then sends updates to server $2$ at rate $\mu_1$.
If server $2$ is busy when a new update arrives, the new update is discarded. Consequently, Server $2$ only accepts updates when it is idle.

The age at the monitor for M/M/1/1
model $\Delta_\text{M/M/1/1}$ can be described by the SHS Markov chain and table of state transitions shown in Fig.~\ref{fig:model-3-shs}. 
The continuous age state vector is $\xv = \rvec{x_0, x_1, x_2}$, where 
$x_0$ is the age of the processed update at the monitor,
$x_1$ and $x_2$ are ages of the update at server $1$ and server $2$ respectively.
For this model, discrete states are $\Qcal = \{0,1\}$, where $0$ and $1$ correspond to server $2$ being idle and busy respectively.
The stationary probabilities are:
 
\begin{equation}\eqnlabel{piq-mm11}
    \pi_0 =  \frac{1}{1+\rho}, \quad\text{and} \quad \pi_1 = \frac{\rho}{1+\rho}. 
\end{equation}

The SHS transitions are self-explanatory. Employing \Thmref{AOI-SHS}, we calculate age at the monitor as $\Delta_\text{M/M/1/1} = v_{0,0} + v_{1,0}$. Some algebraic manipulation gives
\begin{equation}
    \Delta_\text{M/M/1/1}(\mu_2, \rho) = \frac{2}{\mu_2}\Bigl(1+\frac{1}{\rho}\Bigr).
    \eqnlabel{mm11-age-mu2rho}
\end{equation}
Server $1$ is always busy with step $1$, $\Nbar_1(\rho)=1$.
Server 2 working on step 2 is only active in state $q=1$. Using \eqnref{piq-mm11},
\begin{equation} \eqnlabel{n2bar-mm11}
    \Nbar_2(\rho) = \sum_{q\in\Qcal}\pi_q(\rho) n_{q,2} = \pi_1 = \frac{\rho}{1+\rho}.
\end{equation}
The power weighted processor activity for M/M/1/1 is
\begin{equation} \eqnlabel{pn-mm11}
    \PNn{M/M/1/1} = \rho^\al \Nbar_1(\rho) + \Nbar_2(\rho) = \rho^\al + \frac{\rho}{1+\rho}
\end{equation}
with \eqnref{mu2-constraint} equivalent to
\begin{equation}
    \mu_2^\al \le \frac{P}{\E{C}^\al{\PNn{M/M/1/1}}}.
\end{equation}

\subsection{Synchronous Sequential Service (SSS)}
In this model, servers work synchronously, meaning server $1$ generates a fresh update after server $2$ finishes step $2$ on previous update. Consequently, processing on source update starts when both servers are idle.
Here, only one server is busy at any instance.  The age analysis for this model can be approached using either the sawtooth waveform method or the SHS method. For consistency with previous analyses, we apply the SHS method to evaluate the AoI at the monitor. 
\begin{figure}[t]
$\setlength{\extrarowheight}{0.5mm}
\begin{array}[b]{cccc}
l & q_l\to q^\prime_l & \laml &  \xv\Amat_l \\ \hline
1 & 1\to2 & \mu_1 & \rvec{x_0, x_1, x_1} \\
2 & 2\to1 & \mu_2 & \rvec{x_2, 0, x_2}
\end{array}$
\qquad
\begin{tikzpicture}[->, >=stealth', auto, semithick, node distance=1.8cm]
\tikzstyle{every state}=[fill=none,draw=black,thick,text=black,scale=0.85]
\node[state] (0)                    {$1$};
\node[state] (1) [right of=0]  {$2$};
\path
(0) edge[bend left=25, above] node{\small $\mu_1$}  (1)
(1) edge[bend left=25, below] node{\small $\mu_2$} (0);
\end{tikzpicture}
\caption{The SHS transition/reset maps and Markov chain for synchronous sequential servers.}
\label{fig:sync-shs}
\end{figure}
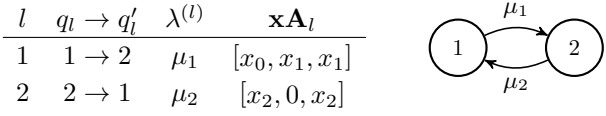

Fig.~\ref{fig:sync-shs} illustrates SHS Markov Chain and table of state transitions for synchronous servers model. The continuous age state vector is $\xv = \rvec{x_0,x_1,x_2}$, where 
$x_0$ is the age of the processed update at the monitor, and 
$x_1$ and $x_2$ are the ages of the update at server $1$ and server $2$ respectively.
For this model, discrete states are $\Qcal = \{1,2\}$, where $1$ and $2$ correspond to server $1$ and server $2$ being busy respectively. We skip explaining the SHS transitions due to space constraints, 
however we do note that unlike in M/M/1/\twostar{} and M/M/1/1 models, $x_1' = 0$ occurs at the transition corresponding to $\mu_2$.
With $\rho = \mu_1/\mu_2$, the Markov Chain in Fig.~\ref{fig:sync-shs} has stationary probabilities
\begin{equation} \eqnlabel{piq-sss}
    \pi_1(\rho) = \frac{1}{1+\rho}, \quad\text{and} \quad \pi_2(\rho) = \frac{\rho}{1+\rho}.
\end{equation}
The age at the monitor, $\Delta_\text{sync}(\mu_1, \mu_2)$, is calculated as $v_{10} + v_{20}$, resulting in
\begin{equation}
    \Delta_\text{SSS}(\mu_2, \rho) = 
    \frac{1}{\mu_2}\Bigl(2+\frac{1}{\rho}+\frac{1}{\rho(1+\rho)}\Bigr).
    \eqnlabel{sss-age-mu2rho}
\end{equation}
Server $1$ works on step 1 only in state $1$ and consequently from \eqnref{piq-sss}
\begin{equation}\eqnlabel{n1bar-sss}
    \Nbar_1(\rho) = \sum_{q\in\Qcal}\pi_q(\rho) n_{q,1} = \pi_1 = \frac{1}{1+\rho}.
\end{equation}
Similarly,  server 2 works on step 2 in state $q=2$. Thus,
\begin{equation}\eqnlabel{n2bar-sss}
    \Nbar_2(\rho) = \sum_{q\in\Qcal}\pi_q(\rho) n_{q,2} = \pi_2 = \frac{\rho}{1+\rho}.
\end{equation}
It follows from \eqnref{pn-def}, \eqnref{n1bar-sss} and \eqnref{n2bar-sss} that
\begin{equation} \eqnlabel{pn-sss}
    \PNn{SSS} = \rho^\al\Nbar_1(\rho) + \Nbar_2(\rho) = \frac{\rho^\al+\rho}{1+\rho}.
\end{equation}
For SSS, \eqnref{mu2-constraint} and \eqnref{pn-sss} imply
\begin{equation}
    \mu_2^\al \le \frac{P}{\E{C}^\al\PNn{SSS}}.
\end{equation}

\section{Problem Formulation: Parallel Servers}
\label{sec:parallel-serv-models}
In this section, we identify two-step update processing models involving two parallel servers, as illustrated in Fig.~\ref{fig:parallel-cartoon} for the case of $n=2$. 
Unlike the series server setup, here each server is allowed to execute both computation steps.
For each model, we identify the system state set $\Qcal$, the stationary probabilities $\pi_q$, age $\Delta(\mu_2, \rho)$, $\PN(\rho)$ and consequently, the upper bound on step 2 service rate $\mu_2$.

We observe that in parallel server systems, the quantity $\rho=\mu_1/\mu_2$ does not always accurately represent the total offered load from step 1 to step 2. This is because one server may transition from step 1 to step 2 while the other remains in step 1.
Moreover, as we will see in certain parallel server policies, even within a single server, the concept of offered load from steps 1 to step 2 can break down, diverging from its conventional interpretation in general queueing theory. 
Therefore, we refrain from referring to $\rho$ as the offered load in parallel setup.  Instead, we treat $\rho$ purely as the ratio between the service rates of step 1 and step 2. 

\subsection{Parallel SSS (P-SSS)}
In this mode, two identical servers process updates independently, in parallel. Each server works on a distinct update and executes both computation steps. 
After processing an update, a server generates new update and immediately starts processing this update.
The total service time for an update is a random variable $T = T_1 + T_2 = C_1/f_1 + C_2/f_2$. With $T_i\sim\exp({\mu_i})$, 
the service time $T$ is a two-parameter hypoexponential distribution with parameters $\mu_1$ and $\mu_2$. 

The age analysis for parallel server setup is non-trivial even with two parallel servers. 
Unlike the series server setup, not every update delivered by the servers resets the monitor's age i.e. not every update delivery is {\em useful}. 
The issue stems from variability in processing times across servers. For instance, one server might take longer to process an update, while the other server, working on a fresher update, finishes its task earlier and resets the monitor's age. Consequently, when the older update from the first server eventually arrives, it fails to reset the monitor's age. 

When server $i, i\in \{1,2\}$, sends an update with age $x_i$ to the monitor, 
the monitor accepts a processed update only if it is fresher than its current update. Consequently, the resulting age at the monitor, denoted by $x_0$, is updated as
\begin{equation} \eqnlabel{psss-x0prime}
x_0' = \min(x_0, x_i).
\end{equation}
Therefore, in the SHS-based age analysis, it is essential to track variables such as $\min(x_0, x_1)$, $\min(x_0, x_2)$, and $\min(x_0, x_1, x_2)$ to accurately account for the acceptance of fresh updates and discarding of outdated ones. This approach to tracking age variables is inspired by the methodology proposed in \cite{Yates-isit2021}.
We now proceed to describe the SHS analysis for the P-SSS model in detail.

\begin{figure}[t]
\centering
\begin{tabular}{c}
\begin{tikzpicture}[->, >=stealth', auto, semithick, node distance=2cm]
\tikzstyle{every state}=[fill=none,draw=black,thick,text=black,scale=1]
\node[state] (1)                    {$1,1$};
\node[state] (2) [right of=1]  {$2,1$};
\node[state] (3) [below of=1] {$1,2$};
\node[state] (4) [below of=2] {$2,2$};
\path
(1) edge[bend left=25, above] node{\small $\mu_1$}  (2)
(1) edge[bend right=25, left] node{\small $\mu_1$}  (3)
(3) edge[bend right=25, below] node{\small $\mu_1$}  (4)
(2) edge[bend left=25, right] node{\small $\mu_1$}  (4)
(2) edge[bend left=25, below] node{\small $\mu_2$}  (1)
(3) edge[bend right=25, right] node{\small $\mu_2$}  (1)
(4) edge[bend right=25, above] node{\small $\mu_2$}  (3)
(4) edge[bend left=25, left] node{\small $\mu_2$}  (2);
\end{tikzpicture} \\
\textbf{(a)} \\
$\setlength{\extrarowheight}{0.5mm}
\begin{array}[b]{cccc}
l & q_l\to q^\prime_l & \laml &  \xv\Amat_l \\ \hline
1 & (1,1)\to(2,1) & \mu_1 & \rvec{x_0, x_1, x_2, x_3, x_4,x_5} \\
2 & (1,1)\to(1,2) & \mu_1 & \rvec{x_0, x_1, x_2, x_3, x_4,x_5} \\
3 & (1,2)\to(2,2) & \mu_1 & \rvec{x_0, x_1, x_2, x_3, x_4,x_5} \\
4 & (2,1)\to(2,2) & \mu_1 & \rvec{x_0, x_1, x_2, x_3, x_4,x_5} \\
5 & (2,1)\to(1,1) & \mu_2 & \rvec{x_3, 0, x_2, 0, x_5, 0} \\
6 & (2,2)\to(1,2) & \mu_2 & \rvec{x_3, 0, x_2, 0, x_5, 0} \\
7 & (1,2)\to(1,1) & \mu_2 & \rvec{x_4, x_1, 0, x_5, 0, 0}  \\
8 & (2,2)\to(2,1) & \mu_2 & \rvec{x_4, x_1, 0, x_5, 0, 0} 
\end{array}$ \\
\textbf{(b)}\\
\end{tabular}
\caption{(a) Markov Chain, and (b) SHS transition maps corresponding to Parallel Sequential Synchronous Service (P-SSS) system.}
\label{fig:parallel-shs}
\end{figure}

The age of processed update for Parallel SSS model can be analyzed using the SHS Markov chain and table of state transitions shown in Fig.~\ref{fig:parallel-shs}. 
The discrete state set is $\Qcal = \{(1,1), (1,2), (2,1), (2,2)\}$, where each tuple $(i,j) \in \Qcal$ represents the current step of server $1$ and server $2$, respectively.
The continuous state age vector is $\xv = \rvec{x_0,x_1,x_2,x_3,x_4,x_5}$, where 
$x_0$ is the age at monitor,
$x_1$ and $x_2$ are the ages of the update at server $1$ and server $2$ respectively, 
$x_3 = \min(x_0, x_1)$, 
$x_4 = \min(x_0, x_2)$, and
$x_5 = \min(x_0, x_1,x_2)$. 

The SHS transitions are enumerated in the table in Fig.~\ref{fig:parallel-shs} and can be understood as follows:
\begin{itemize}
    \item {$l=1,2,3,4$:} In these transitions, the servers only change steps; one of the servers finishes step 1 and begins step 2. Consequently, there is no reset in the age of updates at servers 1 and 2. Since no update is delivered to the monitor, the age at the monitor remains unchanged. As a result, the age variables $x_3$, $x_4$, and $x_5$ also remain unchanged.
    \item {$l=5,6$:} These transitions occur when server 1 finishes processing and delivers the update to the monitor. Server $1$ generates a fresh update, consequently, $x_1' = 0$. Since server 2 continues to work on its update, $x_2' = x_2$. The age at the monitor is reset according to $\min(x_0, x_1)$, which is tracked by age variable $x_3$. Hence, $x_0' = x_3$. Since $x_1' = 0$, thus $x_3' = 0$. The transition for age variable $x_4$ is more complex. 
    We have $x_4' = \min(x_0', x_2') = \min(x_3, x_2) = \min(\min(x_0, x_1), x_2) = \min(x_0, x_1, x_2) = x_5$.
    Further, $x_5' = \min(x_0', x_1', x_2') = \min(x_3, 0, x_2) = 0$.
    \item {$l=7,8$:} These transitions occur when server 2 finishes processing and delivers the update to the monitor. Now server $2$ generates a new update, thus $x_2' = 0$. Since server 1 continues to work on its update, $x_1' = x_1$. The age at the monitor is reset according to $\min(x_0, x_2)$, which is tracked by age variable $x_4$. Hence, $x_0' = x_4$. Since $x_2' = 0$, $x_4' = 0$. 
    Next, we have $x_3' = \min(x_0', x_1') = \min(x_4, x_1) = \min(\min(x_0, x_2), x_1) = \min(x_0, x_1, x_2) = x_5$.
    Additionally, $x_5' = \min(x_0', x_1', x_2') = \min(x_4, x_1, 0) = 0$.
\end{itemize}
The Markov chain in Fig.~\ref{fig:parallel-shs} has stationary probabilities $\piv$
\begin{IEEEeqnarray}{c} \eqnlabel{piq-psss}
    \piv =  \rvec{\pi_{(1,1)},\pi_{(1,2)},\pi_{(2,1)},\pi_{(2,2)}}
    = \frac{1}{(1+\rho)^2} \rvec{1,\rho ,\rho ,\rho^2}.
    \IEEEeqnarraynumspace
\end{IEEEeqnarray}
The age balance equations are
\begin{subequations}
\eqnlabel{parallel-shs-eqn}
\begin{align}
2\rho \mu_2 \vvbar_{1,1} &= \onev[]\pibar_{1,1} + \mu_2\vvbar_{2,1}\Amat_5 + \mu_2\vvbar_{1,2}\Amat_7, \\
(1+\rho)\mu_2 \vvbar_{2,1} &= \onev[]\pibar_{2,1} + \rho\mu_2 \vvbar_{1,1}\Amat_1 + \mu_2\vvbar_{2,2}\Amat_8, \\
(1+\rho)\mu_2\vvbar_{1,2} &= \onev[]\pibar_{1,2} + \rho\mu_2\vvbar_{1,1}\Amat_2 + \mu_2\vvbar_{2,2}\Amat_6, \\
2 \mu_2 \vvbar_{2,2} &= \onev[]\pibar_{2,2} + \rho\mu_2\vvbar_{1,2}\Amat_4 + \rho\mu_2\vvbar_{1,2}\Amat_3.
\end{align}
\end{subequations}
It follows that the age at the monitor can be calculated as $\E{x_3} = v_{(1,1),0} + v_{(2,1),0} + v_{(1,2),0} + v_{(2,2),0}$. Thus,
\begin{align}
    & \Delta_\text{P-SSS}(\mu_2,\rho) \nn
    &\quad= \frac{1}{\mu_2}\biggl(1 + \frac{1}{\rho} + \frac{1+\rho+\rho^2}{4\rho(1+\rho)} + \frac{\rho(1+2\rho)(2+\rho)}{4(1+\rho)^5}\biggr).\eqnlabel{psss-age-mu2rho}
\end{align}

Further, observe that as $\rho \to \infty$
$\Delta_\text{P-SSS} \to 1.25/\mu_2$. This result aligns with the following intuition: 
When $\rho \to \infty$ and $\mu_2$ is finite 
, step $1$ is almost instantly completed, and 
each server delivers an update with an average age $1/\mu_2$. If there were only one server, this would correspond to an average age of $2/\mu_2$ at the monitor, since each update is delivered on average after a duration of $1/\mu_2$.
However, if a single server were running step $2$ at twice the speed ($2\mu_2$), the age at the monitor would be $1/\mu_2$ as 
$\rho \to \infty$. 
However, in the P-SSS setup, we are not running one server at double speed but rather operating two parallel servers. Each server processes an update that is slightly older, resulting in an average age of $1.25/\mu_2$ rather than $1/\mu_2$. 

Next, we observe that in states $(1,2)$ and $(2,1)$, one server works on step 1 while the other works on step 2. In state $(1,1)$, both servers work on step 1, and in state $(2,2)$, no servers work on step 1, as both are executing step 2. Using \eqnref{piq-psss}, the average number of processors working on step 1 is given by
\begin{equation} \eqnlabel{n1bar-psss}
    \Nbar_1(\rho) 
= 2\pi_{(1,1)} + \pi_{(1,2)} + \pi_{(2,1)} = \frac{2}{1+\rho}. 
\end{equation} 
Similarly, the average number of processors working on step 2 is
\begin{equation} \eqnlabel{n2bar-pss}
    \Nbar_2(\rho) 
= \pi_{(1,2)} + \pi_{(2,1)} + 2\pi_{(2,2)} = \frac{2\rho}{1+\rho}.
\end{equation}
Hence, $\PN(\rho)$ for P-SSS is 
\begin{equation} \eqnlabel{pn-psss}
    \PNn{P-SSS} = \rho^\al \Nbar_1(\rho) + \Nbar_2(\rho) = \frac{2(\rho^\al+\rho)}{1+\rho}.
\end{equation}
With $\PNn{P-SSS}$ defined in \eqnref{pn-psss}, the upper bound on the service rate is then
\begin{equation}
    \mu_2^\al \le \frac{P}{\E{C}^\al\PNn{P-SSS}}.
\end{equation}

\subsection{Parallel Coordinated Alternating Freshness (P-CAF)}
When both servers $i$ and $j$ are in step 1, they process the same fresh update concurrently.
If server $i$ transitions to step 2, then server $j$ restarts step 1 with a fresh update. If server $j$ reaches step 2 with its fresher update before server $i$ completes its processing, then server $i$ will abort its current task and restart in step $1$ with a fresh update. 
In this policy, the update in step 1 is always the freshest and only one server is allowed to work on step 2 of update processing at a time. 

\begin{figure}[t]
$\setlength{\extrarowheight}{0.5mm}
\begin{array}[b]{cccc}
l & q_l\to q^\prime_l & \laml &  \xv\Amat_l \\ \hline
1 & 1\to2 & 2\mu_1 & \rvec{x_0, 0, x_1} \\
2 & 2\to2 & \mu_1 & \rvec{x_0, 0, x_1} \\
3 & 2\to1 & \mu_2 & \rvec{x_2, 0,x_2}
\end{array}$
\hfill
\begin{tikzpicture}[->, >=stealth', auto, semithick, node distance=1.8cm]
\tikzstyle{every state}=[fill=none,draw=black,thick,text=black,scale=0.85]
\node[state] (0)                    {$1$};
\node[state] (1) [right of=0]  {$2$};
\path
(0) edge[bend left=25, above] node{\small $2\mu_1$}  (1)
(1) edge[bend left=25, below] node{\small $\mu_2$} (0)
(1) edge[loop right] node{\small $\mu_1$} (1);
\end{tikzpicture}
\caption{The SHS transition/reset maps and Markov chain corresponding to Parallel Coordinated Alternating Freshness (P-CAF) policy for parallel servers.}
\label{fig:cafpolicy-shs}
\end{figure}
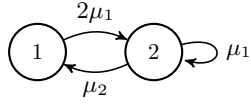
The Markov state space is defined as $\Qcal  = \{1,2\}$, where state 1 corresponds to both servers working step 1, while state 2 indicates that one server is in step 2\ and the other in step 1. The continuous age vector is $\xv = \rvec{x_0, x_1, x_2}$, where 
$x_0$ corresponds to age at the monitor, and
$x_1$ denotes the age of update currently in step 1, and $x_2$ denotes the age of update being processed in step 2.

Note that, in contrast to the P-SSS model where both servers work independently and may deliver stale updates, the P-CAF policy allows coordination among the servers to ensure that only the freshest update that has completed two steps of processing is delivered. 
Mathematically, the analysis is simplified since we don't need to track age variables such as $\min(x_0, x_i)$ as described in \eqnref{psss-x0prime}.  

The SHS Markov chain and table of state transitions are shown in Fig.~\ref{fig:cafpolicy-shs}. 
\begin{itemize}
    \item {$l=1$}: Transition from state 1 to state 2 at rate $2 \mu_1$. In state 1, both servers are in step 1. The time until one server finishes step 1 is the minimum of two independent exponential distributions with rate $\mu_1$, resulting in a departure rate of $2\mu_1$ from state 1.
    Upon transition, one server moves to step 2, so $x_2' = x_1$, while the other server restarts in step 1 with a fresh update, thus $x_1' = 0$.
    The age at the monitor, $x_0$, remains unchanged since no update is delivered, so $x_0' = x_0$.

    \item {$l=2$}: Server in step 1 finishes service to reach step 2, and the server in step 2 (with an older update) restarts in step 1 with a fresh update.
    The age of the update in step 1 is reset to 0, so $x_1' = 0$.
    The age of the update now being processed in step 2 is updated to the age of the previous update in step 1, so $x_2' = x_1$.
    The age at the monitor, $x_0$, remains unchanged, so $x_0' = x_0$.

    \item {$l=3$}: Server in step 2 finishes service and delivers the processed update to the monitor.
    The age at the monitor is updated to the age of the update that was in step 2, so $x_0' = x_2$.
    The server that finished in step 2 and 
    the server that was in step 1 restart with a fresh update, thus $x_1' = 0$.
\end{itemize}
The Markov Chain in Fig.~\ref{fig:cafpolicy-shs} has stationary probabilities
\begin{equation}
    \pi_1 = \frac{1}{1+\rho}, \quad\text{and} \quad \pi_2 = \frac{\rho}{1+\rho}. \eqnlabel{piq-pcaf}
\end{equation}
The age balance equations are
\begin{subequations} \eqnlabel{afpolicy-shs-eqn}
\begin{IEEEeqnarray}{rCl}
2\rho\mu_2\rvec{v_{10}&v_{11}&v_{12}} &=& \rvec{\pi_1&\pi_1&\pi_1} + \mu_2 \rvec{v_{22}&0 &v_{22}},\IEEEeqnarraynumspace \\
(1+\rho)\mu_2 \rvec{v_{20}&v_{21}&v_{22}} &=&\rvec{\pi_2&\pi_2&\pi_2} + \rho\mu_2\rvec{v_{20}&0&v_{21}} \nn
    &&\qquad\qquad + 2\rho\mu_2\rvec{v_{10}&0&v_{11}}.
\end{IEEEeqnarray}
\end{subequations}
Solving the set of equations in \eqnref{afpolicy-shs-eqn}, we can obtain $v_{qj}$. The age at the monitor is $\Delta_\text{P-CAF} = v_{10} + v_{20}$, which yields
\begin{equation}
    \Delta_\text{P-CAF} = \frac{1}{\mu_2}\Bigl(\frac{3}{2(1+\rho)} +\frac{2\rho}{1+2\rho} + \frac{1+\rho+\rho^2}{\rho(1+\rho)^2}\Bigr).
    \eqnlabel{pcaf-age-mu2rho}
\end{equation}
In the P-CAF policy, 
in state $q=1$, both servers execute step 1, 
while in state $q=2$, only one server works on step 1.
Using $\pi_q$ from \eqnref{piq-pcaf}, the average number of servers executing step 1 is
\begin{equation} \eqnlabel{n1bar-pcaf}
     \Nbar_1(\rho) = \sum_{q\in\Qcal} \pi_q(\rho) n_{q,1} = 2\pi_1 + \pi_2 = \frac{2 + \rho}{1+\rho}.
\end{equation}
Since there is only one state ($q=2$) where a server is executing step 2, we have
\begin{equation} \eqnlabel{n2bar-pcaf}
    \Nbar_2(\rho) = \sum_{q\in\Qcal} \pi_q(\rho) n_{q,2} = \pi_2 = \frac{\rho}{1+\rho}.
\end{equation}
It follows from \eqnref{pn-def}, \eqnref{n1bar-pcaf} and \eqnref{n2bar-pcaf} that
\begin{equation} \eqnlabel{pn-pcaf}
    \PNn{P-CAF} = \rho^\al\Nbar_1(\rho) + \Nbar_2(\rho) = \frac{\rho^\al (2+\rho)}{1+\rho} + \frac{\rho}{1+\rho}.
\end{equation}
The upper bound on $\mu_2$ is then given by
\begin{equation}
    \mu_2^\al \le \frac{P}{\E{C}^\al\PNn{P-CAF}}.
\end{equation}

\subsection{Parallel Shared Intermediate Updates (P-SIU)}
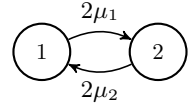
\begin{figure}[t]
$\setlength{\extrarowheight}{0.5mm}
\begin{array}[b]{cccc}
l & q_l\to q^\prime_l & \laml &  \xv\Amat_l \\ \hline
1 & 1\to2 & 2\mu_1 & \rvec{x_0, x_1} \\
2 & 2\to1 & 2\mu_2 & \rvec{x_1, 0}
\end{array}$
\hspace{2.5em}
\begin{tikzpicture}[->, >=stealth', auto, semithick, node distance=1.8cm]
\tikzstyle{every state}=[fill=none,draw=black,thick,text=black,scale=0.85]
\node[state] (0)                    {$1$};
\node[state] (1) [right of=0]  {$2$};
\path
(0) edge[bend left=25, above] node{\small $2\mu_1$}  (1)
(1) edge[bend left=25, below] node{\small $2\mu_2$} (0);
\end{tikzpicture}
\caption{The SHS transition/reset maps and Markov chain corresponding to  Parallel Shared Intermediate Update (P-SIU) policy for parallel servers.}
\label{fig:siupolicy-shs}
\end{figure}
This policy leverages server-to-server communication to share intermediate processing results, enabling parallel execution of each step.
Under this policy, both servers initially work on step 1 of the same update. Once one server completes step 1, it shares the intermediate result with the other server. Subsequently, both servers begin step 2 processing on this intermediate update simultaneously. When either server completes step 2, both servers reset and start processing a fresh update.

The P-SIU policy effectively transforms the parallel server system into a system that behaves like a single server. In this equivalent single-server system, the service times for step $i$ are exponential with rate $2\mu_i$. This simplification arises because the policy ensures that both servers are always working in parallel on the same update, whether in step 1 or step 2.

The discrete state space of the system is defined as $\Qcal = \{1, 2\}$, where state 1 corresponds to both servers in step 1, and state 2 correspond to both servers in step 2. The continuous age vector is $\xv = \rvec{x_0, x_1}$, where 
$x_0$ is the age of the update at the monitor, and
$x_1$ is the age of the update being processed by the servers. 
The SHS Markov Chain and transitions are illustrated in Fig.~\ref{fig:siupolicy-shs} and are self-explanatory. 
Additionally, the Markov Chain in Fig.~\ref{fig:siupolicy-shs} has stationary probabilities
\begin{equation} \eqnlabel{parlsiu-probs}
    \pi_1 = \frac{1}{1+\rho}, \quad\text{and} \quad \pi_2 = \frac{\rho}{1+\rho}.
\end{equation}
The age at the monitor is $\Delta_\text{P-SIU} = v_{10} + v_{20}$, which is expressed as
\begin{equation}
    \Delta_\text{P-SIU} = \frac{1}{\mu_2}\Bigl[1 + \frac{1}{2\rho} \Bigr(1+\frac{1}{1+\rho}\Bigl)\Bigr].
    \eqnlabel{psiu-age-mu2rho}
\end{equation}
In state 1, both servers execute step 1, while in state 2, both servers execute step 2. The average number of servers executing step 1 and step 2 is then
\begin{equation} \eqnlabel{n1bar-psiu}
    \Nbar_1(\rho) = \sum_{q\in\Qcal} \pi_q(\rho) n_{q,1} = 2\pi_1  = \frac{2}{1+\rho}, 
\end{equation}
and
\begin{equation} \eqnlabel{n2bar-psiu}
     \Nbar_2(\rho) = \sum_{q\in\Qcal} \pi_q(\rho) n_{q,2} = 2\pi_2  = \frac{2\rho}{1+\rho},
\end{equation}
From \eqnref{pn-def}, \eqnref{n1bar-psiu} and \eqnref{n2bar-psiu}, we derive the $\PN(\rho)$ for P-SIU as
\begin{equation} \eqnlabel{pn-psiu}
    \PNn{P-SIU} = \rho^\al\Nbar_1(\rho) + \Nbar_2(\rho) = \frac{2(\rho^\al+\rho)}{1+\rho},
\end{equation}
which gives us the upper bound
\begin{equation} \eqnlabel{parlsiu-const}
    \mu_2 \le \frac{P}{\E{C}^\al\PNn{P-SIU}}.
\end{equation}

\section{Numerical Evaluation}
\label{sec:num-eval}
\begin{figure}
    \centering
    \includegraphics{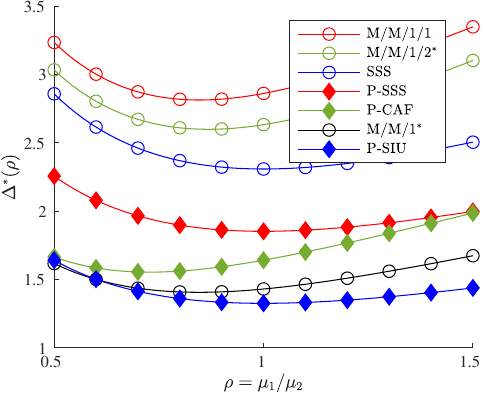}
    \caption{Plot of objective function $\Delta^*(\rho)$ of unconstrained optimization as a function of $\rho$. Here $P=8, \E{C}=1$, and $\alpha=5$. 
    The marker $\circ$ represents servers in series, while $\text{\ding{117}}$ represents parallel servers.
    }
    \label{fig:rhovsfrho}
\end{figure}
In this section, we address the optimization problem presented in \eqnref{opt-problem} for systems identified in Section~\ref{sec:seq-servers-models} and Section \ref{sec:parallel-serv-models}. 
A key observation here is that all the systems we examine end up having a formulation where the age metric is proportional to $1/\mu_2$ times some factor in terms of $\rho$.
This observation is consistent across the models, as reflected in \eqnref{mm1s-age-mu2rho}, \eqnref{mm12s-age-mu2rho}, \eqnref{mm11-age-mu2rho}, \eqnref{sss-age-mu2rho}, \eqnref{psss-age-mu2rho}, \eqnref{pcaf-age-mu2rho}, and \eqnref{psiu-age-mu2rho}.
Thus, in the context of optimization problem \eqnref{opt-problem}, age in all the models is minimized by maximizing $\mu_2$ within the limits imposed by the right side of power constraint \eqnref{mu2-constraint}.
Specifically, for a given power budget $P$, expected CPU cycles required for each step $\E{C}$, and scaling parameter $\al$, denote 
\begin{equation} \eqnlabel{p2-def}
    \PC = \frac{P}{\E{C}^\al \PN(\rho)}.
\end{equation}
Then it follows from \eqnref{mu2-constraint} and \eqnref{p2-def} that the optimal service rate for step 2 is
\begin{equation} \eqnlabel{mu2-opt}
    \mu_2^* = (\PC)^{1/\al}.
\end{equation}
Using \eqnref{mu2-opt}, the constrained optimization problem \eqnref{opt-problem} can be reformulated as an unconstrained optimization problem with the objective function:
\begin{equation}
    \Delta^*(\rho) = \Delta(\mu_2^*, \rho).
\end{equation}


To illustrate the methodology, we begin with a detailed analysis of the \MMonestar{} system. 
Further, to avoid redundancy, we do not present explicit analyses for each model, and instead present the results of the numerical evaluation, which have been derived using the same methodology.
To solve the \MMonestar, we use \eqnref{mm1s-age-mu2rho}, \eqnref{pn-mm1s} and \eqnref{mu2-const-mm1s}. With $\alpha = 5$, the optimization problem in \eqnref{opt-problem} can then be reformulated as:
\begin{subequations} \eqnlabel{mm1s-opt-problem-mu2rho}
\begin{alignat}{2}
    & \text{minimize} &\quad& \frac{1}{\mu_2}\left(1+\frac{1}{\rho}\right) \eqnlabel{mm1s-obj}\\
    & \text{subject to} &\quad& \mu_2^5 \le 
    \frac{P}{\E{C}^5 \Bigl(\rho^5 + \frac{\rho}{1+\rho}\Bigr)}
    , \eqnlabel{mm1s-mu2-constraint} \\
    &&& \mu_2 \ge 0, \text{and } \rho \ge 0.
\end{alignat}
\end{subequations}
We aim to solve \eqnref{mm1s-opt-problem-mu2rho} with respect to the variable $\rho$. Substituting the $\mu_2$ upper bound \eqnref{mm1s-mu2-constraint} for $\mu_2$ in the objective function \eqnref{mm1s-obj}, our goal is then to minimize
\begin{equation}
    \Delta^*_\text{\MMonestar}(\rho) = \frac{\E{C}}{P^{1/5}}\left(\rho^5  + \frac{\rho}{1+\rho}\right)^{1/5} \left(1+\frac{1}{\rho}\right).
    \eqnlabel{mm1s-rhofunc}
\end{equation}
Setting 
$d\Delta^*_\text{\MMonestar}(\rho)/d\rho = 0$, we obtain
$\rho^5(1+\rho) = 4/5$,
yielding the optimal $\rho^* = 0.846$.
Now $\rho^* < 1$,  
implies that server $2$ should operate faster than server $1$, as a slow server 2 would become  a bottleneck in update processing.

Figure \ref{fig:rhovsfrho} illustrates  the objective function $\Delta^*(\rho)$ as a function of $\rho$ for each update processing model.
We first observe, for all systems, that the optimal $\rho^* \le 1$, indicating that step~2 should be processed faster than step~1. For instance, faster step~2 processing in M/M/1/\twostar{} suggests that the queue at server~2 is cleared quickly, which is favorable for minimizing the age. 
The optimal $\rho^*$ for M/M/1/1 is the same as that for \MMonestar{} because, as shown in \eqnref{pn-mm1s} and \eqnref{pn-mm11}, $\PNn{\MMonestar} = \PNn{M/M/1/1}$. Additionally, the age $\Delta_\text{M/M/1/1}(\mu_2, \rho) =2 \Delta_\text{\MMonestar}(\mu_2, \rho)$, as evident from \eqnref{mm11-age-mu2rho} and \eqnref{mm1s-age-mu2rho}. 

For the SSS, P-SSS and P-SIU models, $\rho^* = 1$ indicates that step 1 and step 2 processing should occur at the same rate. This is intuitive due to symmetry: if step 1 is slower, it delays step 2, while if step 2 is slower, the system waits longer to generate a new update. Both scenarios are suboptimal for minimizing the age. 

We observe that, across all considered models, the optimal $\rho^*$ is independent of the power constraint $P$. 
The illustrative example of \MMonestar{} mathematically justifies this, as minimizing the objective function \eqnref{mm1s-rhofunc} will be independent of $P$.
A more conceptual reasoning is as as follows. 
In this work, we have considered a restrictive class of systems where increasing $\mu_2$ and $\mu_1 = \rho\mu_2$  improves age performance. 
In the examined systems, when $\rho$ is fixed, then increasing the service rate at server $2$ is always age reducing as is evident from \eqnref{mm1s-age-mu2rho}, \eqnref{mm12s-age-mu2rho}, \eqnref{mm11-age-mu2rho} \eqnref{sss-age-mu2rho}, \eqnref{psss-age-mu2rho}, \eqnref{pcaf-age-mu2rho} and \eqnref{psiu-age-mu2rho}. Therefore, the optimal  $\mu_2$ should be as large as possible while ensuring that the energy consumed by servers 1 and 2 satisfies the power constraint . 

We note that this independence of $\rho^*$ from $P$ might not hold for all systems. For instance, consider a system where server $1$ generates at will with zero wait and serves at rate $\mu_1$, and updates are queued at server $2$. 
The performance of server 2 is known if the updates arrive fresh \cite{Kaul-YG-infocom2012}. However, in our case, updates arrive with some age from server 1. A longer inter-arrival time between updates can slightly empty the queue at server 2, but the updates arrive with higher age, as the inter-arrival times reflect the age of the updates. 
Hence, it is not straightforward to say that increasing  $\mu_2$ and $\mu_1 = \rho \mu_2$ will always minimize age, and as such there could be some optimal service rates ratio $\rho^*$ which could depend on the power budget $P$.

\begin{figure}[t]
    \centering    \includegraphics{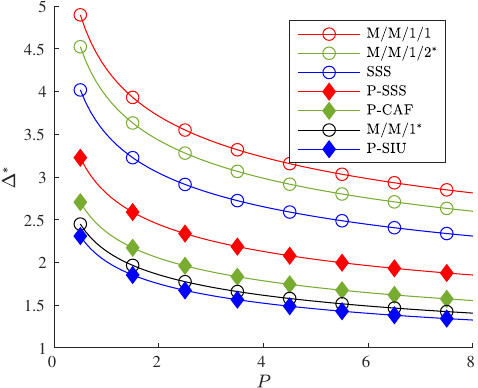}
    \caption{Optimal age 
    $\Delta^* = \Delta(\mu_2^*, \rho^*)$ for servers in series and parallel setups
    under power constraint $P$. Here, $\al=5$ and, $\E{C} = 1$. 
    The marker $\circ$ represents servers in series, while $\text{\ding{117}}$ represents parallel servers.
    }
    \label{fig:seqserv-pvsage}
\end{figure}
Fig.~\ref{fig:seqserv-pvsage} numerically compares optimal age performance $\Delta^* = \Delta(\mu_2^*, \rho^*)$ of all the models as a function of power constraint $P$. As expected, increasing $P$ leads to a larger optimal $\mu_2^*$ resulting in a decrease in age due to the faster service rate. 
It is apparent from Fig.~\ref{fig:seqserv-pvsage} that preemption in service yields better age performance among all servers in series models, which aligns with the existing view in the AoI literature that preemption of old updates by new ones is generally beneficial.
An interesting and somewhat surprising finding is that synchronous service at servers performs better than asynchronous service, indicating that having a single update in service is more advantageous than having multiple updates in progress.
This observation makes sense upon further reflection: synchronous servers prevent updates from lingering in the waiting queue at server $2$ (as in the M/M/1/\twostar{} model), or causing server $2$ to be idle more frequently which occurs when updates are frequently discarded (as in the M/M/1/1 model).

Fig.~\ref{fig:seqserv-pvsage} demonstrates that the P-SSS system achieves better age performance compared to the SSS system, highlighting the advantages of parallel processing over serial processing. P-SSS can be viewed as a parallel version of SSS, where independent servers operate in a manner similar to SSS but with each server consuming power $P/2$, half of the total power budget. Additionally, the superior performance of P-CAF and P-SIU compared to P-SSS further underscores the benefits of incorporating additional information about the state of the other server.

\section{Open Problems: A discussion}
\label{sec:open-probs}

An important and seemingly simple question emerging from this work is: {\em When is power wasted?}
In parallel systems, useless work always wastes power. Specifically, if an update being processed is older than the age at the monitor,
the work is considered useless. If the server continues this useless work, its efforts are deemed wasted.
The P-SSS setup demonstrates an example of wasted effort. In this setup, the two servers work independently on processing updates, without knowledge of the other’s progress. If a server working on an older update had information that the other server had already delivered a fresher update, it would refrain from continuing its work, recognizing that its efforts are redundant. 

In contrast, the P-SIU and P-CAF policies appear to avoid any wasted power. Both policies involve sharing information about the state of the other server, ensuring that no useless work is being done. 
In policies like P-CAF, a server may abandon processing its current update to restart with a new one because the other server was the first to reach step 2. While this update discarding might initially seem wasteful, we argue that it is not wasted since the expected age will be less at the monitor. 

Power can also be wasted even if the work doesn't meet the definition of useless.
In series server setups, the effort of server 1 is wasted if server 2 continues processing an older update despite the availability of a fresher update from server 1. This wasted power is evident in the M/M/1/1 and M/M/1/\twostar{} models. In M/M/1/1, if server 2 is busy when server 1 completes processing, the fresher update from server 1 is discarded, wasting its effort while server 2 continues with its current update. In M/M/1/\twostar{}, server 1 may deliver an update that gets queued at server 2 but is later preempted and discarded by a newer arrival from server 1, as server 2 chooses to continue processing an older update. 

In the \MMonestar{} model, wasted effort is avoided as server 2 always prioritizes the update from server 1. If server 2 is busy when a new update arrives, it preempts its current task to process the fresher update and such preemption is not considered a waste of server 2's effort as it contributes to overall age reduction.

However, we acknowledge that the definition of wasted power becomes less clear in systems with non-exponential service times. For instance, if service times follow a uniform distribution over the interval $[a, b]$, the hazard function  $h(t) = 1/(b-t)$ increases as $t$ approaches $b$, reflecting that the likelihood of update finishing service becomes increasingly certain as time nears the upper bound $b$. Preempting or discarding the update at this stage would not be prudent, as the update is nearly complete and could reset the monitor's age. In such cases, discarding the update would indeed seem like a waste of effort.

On the other hand, there are significant opportunities for age optimization. 
Our analysis assumes a generate-at-will scenario with zero-wait at servers. 
In the existing literature on optimal waiting strategies \cite{Yates-isit2015, Sun-UBYKS-infocom2016UpdateorWait}, it is typically assumed that there is just a single update in the service facility. Upon delivery of this update, the decision to wait is then considered. 
However, in our system, we allow multiple updates to be in process simultaneously. Consequently, the optimality of the known non-zero wait strategies, such as setting a threshold based on prior service time, is unresolved. 


Another approach to age optimization is studying an online policy where service rates are dynamically adjusted based on the system's state. For example in P-CAF system, when one server transitions to step 2, should the other server, now working on a fresher update in step 1, increase its service rate? Alternatively, if the age at the monitor exceeds a certain threshold, the servers could speed up their processing (i.e., age-based service rates). Such adaptive strategies could be studied using a Markov Decision Process (MDP) framework.

Moreover, this work has a natural extension where each processing step has a general service time distribution. The current SHS methodology has a limitation of being applicable to systems with memoryless regimes. Developing a novel SHS analysis for a general service time will not only be useful to this work, but in general to the AoI community.

\section{Conclusion}
This work explored the timely processing of updates that
require a sequence of computational steps. We specifically
identified various  parallel and series server models for update processing, with a focus on understanding the age-power trade-off in the special case of two-step
update processing.
To achieve this, we formulated and solved
optimization problems that determine the optimal service rates
for each step, constrained by a total power budget, to minimize
the average age. 
The analysis revealed that step 2 should generally be faster than step 1 for optimal power efficiency and minimum age.
We also observed that processing by parallel servers has better age performance than servers in series. 

\bibliographystyle{IEEEtran}
\bibliography{output-aoi, output-refs}


\end{document}